\documentclass[prc,aps,fleqn,twocolumn,twoside]{revtex4}

\usepackage{graphicx}
\usepackage{dcolumn}
\usepackage{bm}
\usepackage{amsmath}

\begin{document}

\preprint{Duke-Kyoto-RBRC,NUC-MINN-03/5-T}

\title{Elliptic flow of resonances at RHIC: probing final state interactions
and the structure of resonances}

\author{C.~Nonaka}
\affiliation{Department of Physics, Duke University, Durham, NC 27708, USA}
\author{M.~Asakawa}
\affiliation{Department of Physics, Kyoto University, Kyoto 606-8502, Japan}
\author{S.~A.~Bass}
\affiliation{Department of Physics, Duke University, Durham, NC 27708, USA}
\affiliation{RIKEN BNL Research Center, Brookhaven National Laboratory, 
        Upton, NY 11973, USA}
\author{R.~J.~Fries}
\affiliation{School of Physics and Astronomy, University of Minnesota, Minneapolis, MN 55455, USA}
\author{B.~M\"uller}
\affiliation{Department of Physics, Duke University, Durham, NC 27708, USA}

\date{\today}

\begin{abstract}
We propose the measurement of the elliptic flow of hadron resonances
at the Relativistic Heavy Ion Collider as a tool to probe the amount
of hadronic final state interactions for resonances at intermediate and large
transverse momenta. This can be achieved by looking at systematic
deviations of the measured flow coefficient $v_2$ from the scaling law
given by the quark recombination formalism.
Our method can be generalized to explore the structure of exotic 
particles, such as the recently found pentaquark $\Theta^+ (1540)$.
\end{abstract}

\pacs{Valid PACS appear here}
\maketitle

Recently, progress has been made
in our understanding of the hadronization of a quark gluon plasma
created in high energy heavy ion collisions. Both the yields and the
anisotropic flow $v_2$ of hadrons at transverse momenta above 2 GeV,
observed at the Relativistic Heavy Ion Collider (RHIC),
are described well by a combination of two mechanisms: (i) the 
statistical recombination of thermal constituent quarks from a quark gluon 
plasma phase during the phase transition, (ii) fragmentation of jets which 
experienced energy loss due to the surrounding medium 
\cite{Duke:02,Duke:03,Duke:04,Texas:02,HwYa}.

The elliptic flow $v_2$ is defined as 
the second Fourier coefficient of the azimuthal hadron
spectrum \cite{v2_original}
\begin{equation}
  \frac{d^2N}{P_TdP_T d\Phi} = \frac{dN}{2\pi P_TdP_T} 
   [ 1 + 2v_2(P_T)\cos 2\Phi + \ldots].
\end{equation}
It measures the elliptic anisotropy in the spectrum  
since
\begin{equation}
  v_2(P_T) = \langle \cos 2 \Phi \rangle 
  = \frac{\int d \Phi \cos 2 \Phi \frac{d^2 N}{P_T d P_T d\Phi}}
  {\int d \Phi  \frac{d^2N}{P_T d P_T d\Phi}}.
\label{eq:v2def}
\end{equation}
Such an asymmetry naturally arises in heavy ion collisions with non-vanishing
impact parameter $b>0$. 

It has been shown that in the region 
where recombination 
of partons dominates the hadronization process --- i.e., \ for $P_T < 4$ $(6)$ 
GeV for mesons (baryons) --- and mass effects are suppressed 
--- i.e., \ $P_T > 2$ GeV --- $v_2$ obeys the simple scaling law,
\begin{equation}
  v_2(P_T) = n v_2^q(P_T/n),
\label{eq:scaling}
\end{equation}
where $n$ is the number of valence quarks of the hadron 
\cite{Duke:03,Vo,Ohio:03}. 
We will assume throughout the paper that strange quarks and light quarks have 
the same elliptic flow $v_2^q$. 
Measurements of strange hadrons support this assumption \cite{Duke:03}. 
At higher transverse momentum, $P_T> 6 $ GeV, where fragmentation of partons
dominates, we expect the $v_2$ of all hadrons to lie on a universal curve since the
effect of fragmentation functions largely cancels in Eq.\ (\ref{eq:v2def}).
We refer the reader to Refs.\ \cite{Duke:03,Duke:04} for a more detailed 
discussion.

The scaling law (\ref{eq:scaling}) has been impressively confirmed by 
measurements at RHIC. Pions and kaons ($n=2$), protons, $\Lambda$ and $\Xi$
($n=3$) 
all fall on one universal curve, if $v_2$ and $P_T$ are divided by the number
$n$ of valence quarks \cite{rhic_exp_v2_strange,rhic_exp_v2}. 
Slight deviations can be observed and
are largest for pions. This can be due to its nature as a Goldstone boson ---
its mass is much smaller than the sum of its constituent quark masses
--- or because
a large fraction of all pions are not created at hadronization but 
by the subsequent decay of hadron resonances \cite{Texas:res}.
This poses the interesting question how the existence of a hadronic phase 
generally affects hadron production at intermediate and large momenta. 
We suggest that the amount of hadronic rescattering can be determined by 
measuring $v_2$ for several hadronic resonances. 


Hadrons can experience rescattering after hadronization, until kinetic 
freeze-out takes place. Experimentally, it has been found that the 
kinetic freeze-out temperature at RHIC \cite{STAR:kinetic} 
is much lower than the expected 
QCD phase transition temperature \cite{MILC:02}.
Resonances produced from a hadronizing quark gluon plasma decay in 
the hadronic medium. If their decay products rescatter, 
the signal is lost  
in the experimental measurement, since the original resonance
cannot be reconstructed from a correlation analysis of its daughter
hadrons. Likewise, hadrons from the medium may scatter into a resonance
state and thus contribute to the final measured yield. The total measured
resonance yield will therefore have two contributions: 
\begin{enumerate}
\item {\it primordial resonances:} resonances produced from a hadronizing
 	quark gluon plasma whose decay products have not rescattered 
 	(QGP mechanism).
\item {\it secondary resonances:} resonances produced in the hadronic 
final state	via hadron-hadron rescattering (HG mechanism).
\end{enumerate}



In this paper, we will show that $v_2$ differs between the two cases.
Key to our analysis is the observation that 
the elliptic flow $v_2$ will be {\it additive} for any type of composite 
particle with respect to the $v_2$ of its constituents. 
Note that this feature is mimicked in a hydrodynamic model when  
mass effects are neglected. Then the elliptic flow of
all particles follows the same line through the origin, $v_2(P_T) \propto  P_T$. 
The scaling law simply maps this line into itself.
Only the deviation from the ideal hydro behavior with increasing $P_T$,
manifest in the saturation of $v_2$ at intermediate $P_T$, makes the
scaling law non-trivial.
The 
$v_2$ of a $K^*_0$ meson in the recombination domain will therefore 
be the sum of the 
$v_2$ contributions of the $d$ and $\bar s$ quarks if it has been formed from 
a hadronizing QGP, or the sum of the contributions of the $K^+$ and $\pi^-$ 
if it has been formed through the coalescence of a kaon and a pion in the
hadron phase. 
Our discussion of deriving the formalism for the coalescence of 
partons can be extended to the coalescence of hadrons in a 
straightforward way \cite{Duke:02, Duke:03, Duke:04}. 
Therefore we can here assume the same additive feature for hadron-hadron 
coalescence as well.
Naively, if the kaon and pion themselves are formed by
quark recombination, we expect a scaling of the $K^*_0$ with $n=4$ in the HG
case, while it scales with $n=2$, like the stable mesons, if it is created
at the phase transition. The scaling will be altered by the possibility that
one or both of the coalescing hadrons could come from fragmentation, which
breaks the scaling law. The final result will be a mixture of both scenarios. 

We assume that the $v_2$ of stable hadrons (stable with
regard to strong interactions) is not affected by hadronic interactions
for $P_T$ above 2 GeV. This is a sound assumption in light of the
experimental facts, as discussed above. 
Let $r(P_T)$ be the fraction of resonances that have escaped from 
hadronization without rescattering of the decay products vs. the total measured
yield.
Given the knowledge of $v_2$ in the two limiting cases, 
the QGP mechanism and the HG mechanism, 
the fraction $r(P_T)$ can be determined by measuring $v_2$ 
at intermediate $P_T$, between 2 and 6 GeV via
\begin{equation}
  v_2^{\rm measured}=r(P_T) v_2^{\rm QGP} + (1-r(P_T)) v_2^{\rm HG}.
\label{resfrac}
\end{equation}

%
%
One may be able to deduce the widths and cross sections of resonances in the 
hadronic medium by measuring $r(P_T)$ for several resonances.  
Our scheme can be 
generalized to explore the structure of exotic particles, e.g. a large 
molecular state, as discussed in the case of the recently discovered 
pentaquark $\Theta^+(1540)$.
We will discuss this in more detail below.



\begin{figure}[thb]
\centering
\includegraphics[scale=0.55]{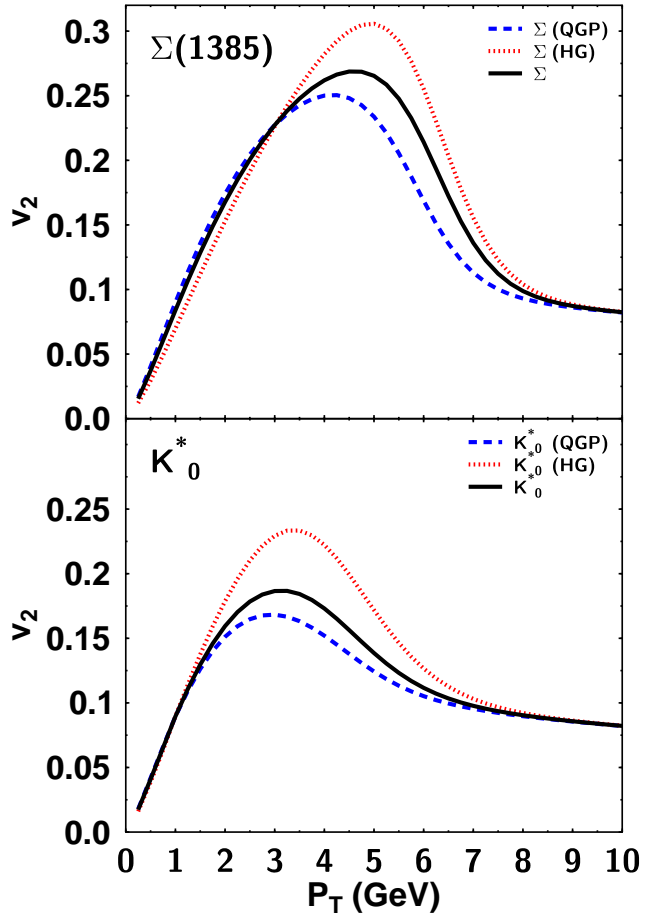}
\caption{(Color online) Elliptic flow $v_2$ as a function of $P_T$ for the $\Sigma^*(1385)$ 
(top) and the $K^*_0(892)$ (bottom). Dashed line: resonance contribution 
from hadronization (QGP); dotted line: resonance contribution from 
coalescence in a hadron gas (HG); solid line: calculation for a 30\% 
contribution of hadronic resonance formation.}
\label{v2_res}
\end{figure}

Figure~\ref{v2_res} shows the elliptic flow $v_2$ as a function of transverse
momentum $P_T$ for the $\Sigma^*(1385)$ (top) and the $K^*_0$ (bottom) 
resonances. The dashed lines refer to primordial 
resonance production from a hadronizing
QGP via parton recombination at low and intermediate $P_T$ and fragmentation
at large $P_T$. The dotted lines refer to these resonances being produced
via hadronic final state interaction through hadron--hadron recombination:
$\Lambda+\pi \to \Sigma^*$ and $K^+ + \pi^- \to K^*_0$. These two
scenarios establish lower and upper limits for $v_2$.
The solid line represents a calculation with contributions from both,
the QGP and the HG mechanisms.
For demonstration purposes we have here chosen $r(P_T)$ to be constant, 
with a value of 70\%. $r(P_T)$ is the quantity that can 
ultimately be extracted from data.

For quark recombination we used the scaling law (\ref{eq:scaling}) together
with the parton $v_2^q$ extracted and discussed in \cite{Duke:03,Duke:04}.
Since the fragmentation
functions for the resonances are unknown, we utilize the universal curve 
for the $v_2$ contribution from fragmentation as obtained in
\cite{Duke:03,Duke:04}.

Due to the transition from the recombination to the fragmentation regime, 
the maxima of the HG and QGP curves do not scale like
5/3 and 4/2 for the $\Sigma^*$ and $K^*_0$ respectively. Instead, the HG curve
is brought down by contributions from pions, kaons or $\Lambda$s from 
fragmentation, violating the scaling law. Still, the difference between the
two limiting cases can be as large as 30\% between 2 and 6 GeV.



The fraction of the total resonance yield being produced by re-generation
in the hadronic phase
cannot be inferred through the final measured resonance yield since 
the amount of signal loss in resonance reconstruction
due to rescattering of the resonance decay products cannot be 
experimentally determined. The scheme outlined above to
determine $r(P_T)$ therefore provides complimentary information 
to the resonance yield measurements via invariant mass reconstruction.
The determination of $r(P_T)$ allows for an estimate of the gain term 
of resonance production in the hadronic phase - together with the 
yield information this allows for a far better estimate of the actual 
QGP production of resonances and for an improved  estimate of the 
signal loss due to rescattering of the decay products.
These estimates are of high relevance 
for the interpretation of statistical model fits with respect 
to resonances yields.

We can now utilize the scaling law for the elliptic flow -- and
its transition to a universal value for $v_2$ in the fragmentation domain --
to probe the structure of exotic particles, such as the $\Theta^+ (1540)$
pentaquark state. The discovery of this novel hadronic state with
five valence quarks ($uudd\bar{s}$) has recently been reported in 
\cite{Nakano:03}. 
Its existence was predicted in 1997 by Diakonov {\it et al.} in the 
chiral quark soliton model (CQSM) \cite{Diakonov:1997mm}.
Several other groups have by now confirmed this important discovery
\cite{Barmin:03,Stephanya:03,Barth:03}.
The $\Theta^+$ is the first unambiguous five quark state,
since other candidates for five quark states, such as
the $\Lambda (1405)$, have quark compositions
that can mix with three quark states like
$udd\bar{d}s \leftrightarrow uds$. Recently, NA49 reported the observation
of two more states that could be the two missing pentaquarks
in the antidecuplet \cite{NA49:03}. The yield of the $\Theta^+$ in
heavy ion collisions has been estimated in \cite{Randrup:03, Texas:03}.

While the discovery of the $\Theta^+$ has undoubtedly established the 
existence of pentaquarks, its structure, and even its spin 
and parity, have not yet been experimentally verified. 
It is still an open question whether 
pentaquarks like the $\Theta^+$ have a compact structure with all 
five valence quarks being closely confined, just as the three quarks 
in the nucleon, or whether they are  molecular bound states of a baryon and a 
meson. A bound state of two diquarks and an antiquark
was recently suggested as well \cite{JaffeWilczek:03}.

Loosely bound states with large geometric cross section whose binding 
energies are much less than the temperature,
like the deuteron, are dissociated in the hadron phase
even if they were produced at hadronization. 
Such a medium induced breakup reaction does not allow for the
reconstruction of the state via two-particle mass spectra, since the initial
momentum of the scattering partner is unknown. These bound states are 
unobservable.
Thus, the loosely bound states and molecular states observed in heavy ion 
collisions are produced close to the kinetic freeze-out 
hypersurface \cite{Texas:03}, in the case of the pentaquark by 
$K^+ + n \rightarrow \Theta^+$ and $K^0 + p \rightarrow \Theta^+$ 
(summarily denoted as $KN$).
The hadronic phase can be viewed as a filter that tends to 
let the $\Theta^+$ from hadronization only survive if it is a compact 
five quark state.

With respect to the azimuthal anisotropy of pentaquarks produced in
heavy-ion collisions at RHIC, we expect it to fulfill the scaling law
(\ref{eq:scaling}) with $n=5$. This would be an impressive manifestation of its
exotic five quark character. On the other hand, we expect modifications to 
hold, as we discussed them for conventional resonances. However, we want
to point out, that in the case of conventional baryon resonances the 
scaling law holds for $n=3$ in the QGP scenario, while it is roughly $n=5$
in the HG case. For a pentaquark it would be $n=5$ in both cases. The 
difference would merely come from the possibility that for the
reaction $N+ K \to \Theta^+$ one or both of the initial hadrons could come 
from fragmentation. The same is true for a deuteron, that scales approximately
with $n=6$ in both cases.

If the $\Theta^+$ is a molecular state, the momentum fraction $x_K$
of the kaon in a lightcone frame is expected to be approximately given by
$r_K=M_K/(M_K + M_N) \sim 0.35$, where $M_K$ and $M_N$ are
the masses of the kaon and nucleon, respectively. We therefore use the wave
functions \cite{Duke:03},
\begin{eqnarray}
  | \phi_{\Theta} (x_K, x_N ) |^2
  & = & \delta \left ( x_K - 0.35 \right )
 , \\
  | \phi_{d} (x_p, x_n ) |^2
  & = & \delta \left ( x_p - 0.5 \right )
 .
\end{eqnarray}



\begin{figure}[thb]
\centering
\includegraphics[scale=0.55]{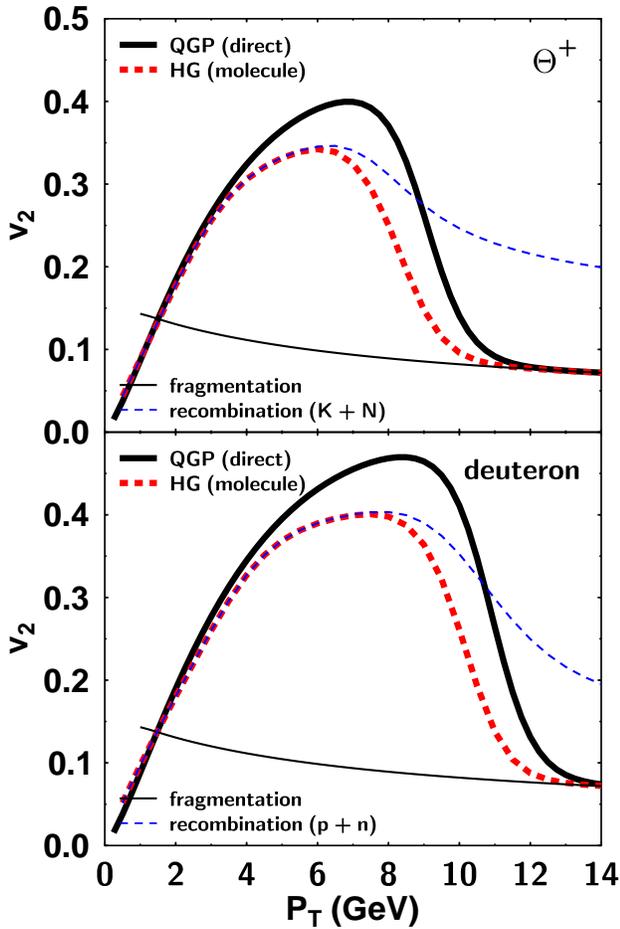}
\caption{(Color online) $P_T$ dependence of $v_2$ for the pentaquark $\Theta^+$ (top)  
and the deuteron $d$. 
Thick solid line: fragmentation and recombination of a 5 (6) quark state 
for $\Theta^+$ ($d$); thick dashed line: fragmentation and recombination of 
$K+N$ ($p+n$) for $\Theta^+$ ($d$) as true for a 
molecular state; thin solid line: direct fragmentation (universal curve); 
thin dashed line: recombination of $K+N$ ($p+n$) only. }
\label{v2scenarios}
\end{figure}



In Fig.\ \ref{v2scenarios} we show $v_2$ for the different scenarios
discussed above. 
Note that the scenario with all pentaquarks from hadronization
(QGP) is now providing the \emph{upper} limit, while it was the lower limit
in the case of conventional resonances.
We include the deuteron in the study to give an example of a well known 
molecular state.
Above  $P_T=3$ GeV we find a considerable difference, up to 20\%,  
between the scenario where the pentaquark is created at the phase 
transition --- and is therefore a compact five quark state
--- and the scenario where it originates from $KN$ coalescence. 
If a $KN$ coalescence takes place to form a $\Theta^+$ between 
4 and 8 GeV, it is likely that either the kaon or neutron (or both) stems 
from fragmentation which lowers the resulting $v_2$.
At very high $P_T$, both the kaon and the nucleon have to
come from fragmentation. This results in a power law spectrum that
 falls off like $P_T^{-2b}$, approximately twice as rapidly as a directly
fragmenting particle with $\sim P_T^{-b}$.
Therefore, direct fragmentation will dominate at very large $P_T$ and will
eventually lead $v_2$ down to the universal curve.

For the deuteron as a shallow molecular bound state we expect $pn$ 
recombination at the kinetic freeze-out to dominate. The deuteron is therefore
ideal as a benchmark measurement. First results for the $v_2$ of the deuteron
have been announced by 
the PHENIX collaborations 
\cite{Sakai:03}.
As expected, this particle seems to follow the scaling law with $n=6$ within
the large error bars. With the upcoming high statistics Au+Au run, error bars
and $P_T$ coverage should be greatly improved, allowing for a gauging of the 
formalism.


Let us comment on the third possible scenario for the $\Theta^+$, that was 
advocated by Jaffe and Wilczek \cite{JaffeWilczek:03}.
If it is a collection of two tightly bound $(ud)$ diquarks and an 
$\bar{s}$ quark, recombination would predict the $P_T$ dependence
of $v_2$ to be very similar to that of a ``democratic'' 5 quark state.
This, however, is only true if the diquarks are also formed at hadronization 
and the individual quarks have not been correlated early on with enough time
to have thermalized diquark states. In that case one would expect
a scaling with $n=3$. 

In summary, we propose the measurement of the elliptic flow of hadronic 
resonances at the Relativistic Heavy Ion Collider. After recombination scaling
has been confirmed for the $v_2$ of stable particles, deviations from 
this scaling for resonances can directly measure the fraction of 
resonances that escape from 
hadronization without rescattering in the hadronic phase at intermediate and 
large transverse momentum. 
This allows conclusions to be drawn about  
the importance of the hadronic phase in this region of phase space.

In turn, this method can be applied to explore the structure of exotic 
particles, such as the recently found $\Theta^+(1540)$.
The amount of rescattering in the hadronic phase is sensitive to the size of
the system, eventually distinguishing between a compact hadron and a 
molecular state. The deuteron, with a known molecular structure, can serve as
a precise benchmark for these measurements.
The same method is also applicable to determine
the structure of other exotic candidates with a possible molecular structure
such as $\Lambda (1405)$, $a_0 (980)$, $f_0 (980)$,
and so on \cite{Jaffe:77,AlfordJaffe:00,Close:02}.


\begin{acknowledgments}
We thank J.\ Kapusta, V. Greco and C.M. Ko for stimulating discussions.
This work was in part supported in part by RIKEN, 
Brookhaven National Laboratory, 
Grant-in-Aid by the Japanese Ministry of Education No. 14540255,
and DOE grants DE-FG02-96ER40945 and DE-AC02-98CH10886.
S.A.B. acknowledges support from an Outstanding Junior Investigator
Award (DOE grant DE-FG02-03ER41239) and R.J.F. has been supported
by DOE grant DE-FG02-87ER40328.
\end{acknowledgments}


\begin{thebibliography}{99}

\bibitem{Duke:02}R.~J.~Fries, B.~M\"uller, C.~Nonaka, and S.~A.~Bass,
Phys. Rev. Lett. {\bf 90}, 202303 (2003). 

\bibitem{Duke:03}R.~J.~Fries, B.~M\"uller, C.~Nonaka, and S.~A.~Bass,
Phys.\ Rev.\ C {\bf 68}, 044902 (2003).

\bibitem{Duke:04}
C.~Nonaka, R.~J.~Fries, and S.~A.~Bass, 
Phys.\ Lett.\ B {\bf 583}, 73 (2003). 

\bibitem{Texas:02}V.~Greco, C.~M.~Ko, and P.~L\'evai,
Phys.\ Rev.\ Lett. {\bf 90}, 202302 (2003); 
V.~Greco, C.~M.~Ko, and P.~L\'evai, 
Phys.\ Rev.\ C {\bf 68}, 034904 (2003).

\bibitem{HwYa}
R.~C.~Hwa and C.~B.~Yang, Phys.\ Rev.\ C {\bf 67}, 034902 (2003). 

\bibitem{v2_original}
J.~Y.~Ollitrault,
Phys.\ Rev.\ D {\bf 46}, 229 (1992).

\bibitem{Vo}
S.~A.~Voloshin, Nucl.\ Phys.\ A {\bf 715}, 379 (2003). 


\bibitem{Ohio:03}
D.~Moln\'ar and  S.~A.~Voloshin, 
Phys.\ Rev.\ Lett.\ {\bf 91}, 092301 (2003);
Z.~Lin and D.~Moln\'ar, Phys.\ Rev.\ C {\bf 68}, 044901 (2003).

\bibitem{rhic_exp_v2_strange}
J.~Castillo [STAR Collaboration], talk at QM2004, Oakland, USA, 
January 11-17, 2004. 

\bibitem{rhic_exp_v2}
P.~Sorensen [for the STAR Collaboration], J.\ Phys.\ G {\bf 30}, S217 (2004).

\bibitem{Texas:res}
V.~Greco and C.~M.~Ko,
nucl-th/0402020.

\bibitem{STAR:kinetic}
J.~Adams {\it et al.} [STAR Collaboration], nucl-ex/0310004.

\bibitem{MILC:02}
C.~Bernard {\it et al.}, Nucl.\ Phys.\  B (Proc. Suppl.) {\bf 119}, 523 (2003).

\bibitem{Nakano:03}
T.~Nakano {\it et al.} [Leps Collaboration], 
Phys. Rev. Lett. {\bf 91}, 012002 (2003).

\bibitem{Diakonov:1997mm}
D.~Diakonov, V.~Petrov, and M.~V.~Polyakov,
Z.\ Phys.\ A {\bf 359}, 305 (1997).

\bibitem{Barmin:03}
V.~V.~Barmin,   
Phys.\ Atom.\ Nucl.\  {\bf 66}, 1715 (2003)
[Yad.\ Fiz.\  {\bf 66}, 1763 (2003)] 
(hep-ex/0304040).

\bibitem{Stephanya:03}
S.~Stepanyan 
{\it et. al.} [CLAS Collaboration], 
Phys.\ Rev.\ Lett.\ 91, 252001 (2003).

\bibitem{Barth:03}
J.~Barth {\it et al.} [SAPHIR Collaboration],
Phys.\ Lett.\ B {\bf 572}, 127 (2003).

\bibitem{NA49:03}
C.~Alt {\it et al.} [NA49 Collaboration],
hep-ex/0310014.

\bibitem{Randrup:03}
J.~Randrup,
Phys.\ Rev.\ C {\bf 68}, 031903 (2003);
J.~Letessier, G.~Torrieri, S.~Steinke and J.~Rafelski, 
hep-ph/0310188.

\bibitem{Texas:03}
L.~W.~Chen, V.~Greco, C.~M.~Ko, S.~H.~Lee, and W.~Liu,
nucl-th/0308006.

\bibitem{JaffeWilczek:03}
R.~L.~Jaffe and F.~Wilczek,
Phys.\ Rev.\ Lett.\ {\bf 91}, 232003 (2003).



\bibitem{Sakai:03}
M.~Kaneta [PHENIX Collaboration], talk at QM2004, Oakland, USA,
January 11-17, 2004.

\bibitem{Jaffe:77}
R.~L.~Jaffe,
Phys.\ Rev.\ D {\bf 15}, 267 (1977); {\it ibid.} {\bf 15}, 281 (1977).

\bibitem{AlfordJaffe:00}
M.~Alford and R.~L.~Jaffe,
Nucl.\ Phys.\ B {\bf 578}, 367 (2000).

\bibitem{Close:02}
F.~E.~Close and N.~A.~T\"ornqvist,
J. Phys. G {\bf 28}, R249 (2002).

\end{thebibliography}
\end{document}